\documentclass[runningheads,a4paper]{llncs}
\usepackage{times}
\usepackage{listings}
\usepackage{colonequals}
\usepackage{xspace}
\usepackage{amssymb}
\usepackage{graphicx}
\usepackage{stmaryrd}
\usepackage{cite}
\usepackage{hyperref}
\usepackage{framed}

\usepackage{xcolor}

\usepackage{amsmath}    
\usepackage{booktabs}

\usepackage{url}
\urldef{\mailsa}\path|firstname.lastname@epfl.ch|

\newcommand{\cvc}{\textsc{cvc}{\small 4}\xspace}

\newcommand{\teq}{\approx}

\newcommand{\M}{\mathcal{M}}

\usepackage{defs}

\usepackage{program}

\begin{document}
\sloppy

\mainmatter  

\newcommand{\mytitle}{Approaches for Synthesis Conjectures in an SMT Solver}
\title{\mytitle}

\titlerunning{\mytitle}

%
%
\author{Andrew Reynolds}


\institute{
{\'E}cole Polytechnique F{\'e}d{\'e}rale de Lausanne (EPFL), Switzerland
\email{\{firstname.lastname\}@epfl.ch}
}


%
%

\maketitle

\begin{abstract}
This report describes several approaches for handling synthesis conjectures within an Satisfiability Modulo Theories (SMT) solver. 
We describe approaches that primarily focus on determining the unsatisfiability of the negated form of synthesis conjectures using new techniques for quantifier instantiation.
\end{abstract}

\section{Synthesis in an SMT solver}

A \emph{synthesis conjecture} states there exists a function $f$ for which some universal property $P$ holds.
In other words, a conjecture of this form can be stated as:
\begin{eqnarray} \label{eqn:syn_conj}
\exists f. \ \forall \vec i. P( f, \vec i )
\end{eqnarray}
where $f$ is a function to synthesize, and $P$ states the property that $f$ must satisfy for all $\vec i$.
In this report, 
we examine approaches for handling conjectures of this form within the core of a Satisfiability Modulo Theories (SMT) solver~\cite{CVC4-CAV-11}.

For determining the satisfiability of the formula~\eqref{eqn:syn_conj}, 
an SMT solver may treat $f$ as an uninterpreted function,
and find an interpretation for $f$ for which $\forall \vec i. P( f, \vec i )$ is satisfied for all $\vec i$.
Notice this task poses multiple challenges to the SMT solver.
First, the solver must construct a stream of candidate interpretations for $f$ based on its partial model, 
which by default gives no guarentee that an interpretation will eventually be discovered that satisifies this (or other) quantified formulas.
Moreover, the solver must be extended with methods for determining when universally quantified formulas are satisfied.
In fact, showing that a universally quantified formula is satisfied for all $\vec i$ often is accomplished by showing that 
its negation under the candidate interpretation of $f$ is unsatisfiable~\cite{GeDeM-CAV-09}, which itself reduces to a ground satisfiability query.
An alternative line of research in the domain of software verification
has explored specialized techniques for establishing the satisfiability of quantified horn clauses~\cite{beyene2013solving, beyene2014constraint, DBLP:conf/sas/BjornerMR13} 
which has had success for handling clauses in several theories.


More traditional designs of SMT solvers for handling quantified formulas have focused on instantiation-based methods
that consider ground instances of quantified formulas until a refutation is found at the ground level~\cite{Detlefs03simplify:a}.
While such techniques are incomplete in general,
it has been shown they are quite effective in practice for finding proofs of unsatisifiability~\cite{DBLP:conf/cade/MouraB07, reynolds14quant_fmcad}.
Arguably, doing so is more natural for an SMT solver and poses fewer complications than establishing the satisfiability of quantified formulas.
For this reason, we advocate approaches for synthesis that instead establish the \emph{unsatisfiability} of the negation of the aforementioned conjecture:
\begin{eqnarray} \label{eqn:neg_syn_conj}
\forall f. \ \exists \vec i. \neg P( f, \vec i )
\end{eqnarray}
This seemingly poses another challenge to the SMT solver, namely, the outermost quantification is second-order, as it quantifies over functions $f$, 
which no SMT solver to our knowledge directly supports.
However, this report presents two techniques for common cases of such conjectures that avoid the need for second-order quantification.
We will first examine the case of \emph{syntax-guided synthesis}, where our problem additionally takes as input a
syntax defining the space of possible solutions.
A recent line of research has targetted such problems~\cite{6679385}, since they are noted to be of practical interest in various applications.

\subsection{Syntax-Guided Synthesis}
\label{sec:syntax-guided}

Consider a negated synthesis conjecture of the form $\forall f. \ \exists \vec i. \neg P( f, \vec i )$.
If our space of solutions for $f$ is restricted to some syntax (that is, a signature of symbols that can be used to construct $f$),
we may consider $f$ to be a variable $g$ of sort $S$, where $S$ is an algebraic datatype 
whose constructors represent the programs that may be used in solutions for $f$.
In this report, the notions of syntax specifications and datatypes will be used interchangably.
Consider the following, which we use as a running example.

\begin{example}
\label{ex:max-sygus}
Consider the following property for which a function $f : Int \times Int \rightarrow Int$ must satisfy,
namely that $f$ computes the maximum of two input integers $x$ and $y$:
\[\begin{array}{l@{\hspace{1em}}l}
P_0 := \lambda fxy. \ f( x, y ) \geq x \wedge f( x, y ) \geq y \wedge ( f( x, y ) \teq x \vee f( x, y ) \teq y )
\end{array}\]

Say our solutions for $f$ are restricted to a syntax $S$, which we may represent as the following inductive datatypes:
\[\begin{array}{l@{\hspace{1em}}l@{\hspace{1em}}l}
S & := & \mathsf{0} \mid \mathsf{1} \mid \mathsf{x} \mid \mathsf{y} \mid S_1 \mathsf{+} S_2 \mid S_1 \mathsf{-} S_2 \mid \mathsf{ite}( C_1, S_1, S_2 ) \\
C & := & S_1 \mathsf{\leq} S_2 \mid S_1 \mathsf{\teq} S_2 \mid C_1 \mathsf{\wedge} C_2 \mid C_1 \mathsf{\vee} C_2 \mid \mathsf{\neg} C_1
\end{array}\]
This defines a term signature that includes variables $x$, $y$, and theory symbols with builtin interpretations known by the SMT solver,
where terms of sort $S$ refer to those of sort $Int$ and terms of sort $C$ refer to those of sort $Bool$.
\footnote{
It is important to note that the symbols shown in the definition of $S$ and $C$ denote datatype constructors, 
and are not to be confused with the builtin theory operators they correspond to.
This will be unambiguous from the context in which we use these symbols.
}

To state properties of terms in this syntax,
we introduce an uninterpreted function for each datatype, which we refer to its \emph{evaluation operator}.
Let $e_S$ be a function of sort $S \times Int \times Int \rightarrow Int$ and $e_C$ be a function of sort $C \times Int \times Int \rightarrow Bool$,
and $\mathcal{A}_S \cup \mathcal{A}_C$ be the axiomatization of these functions respectively, containing:
\[\begin{array}{l@{\hspace{1em}}l}
\forall xy. \ e_S( \mathsf{0}, x, y ) \teq 0 & \forall s_1 s_2 xy. \ e_C( s_1 \mathsf{\leq} s_2, x, y ) \teq (e_S( s_1, x, y ) \leq e_S( s_2, x, y )) \\
\forall xy. \ e_S( \mathsf{1}, x, y ) \teq 1 & \forall s_1 s_2 xy. \ e_C( s_1 \mathsf{\teq} s_2, x, y ) \teq (e_S( s_1, x, y ) \teq e_S( s_2, x, y )) \\
\forall xy. \ e_S( \mathsf{x}, x, y ) \teq x & \forall c_1 c_2 xy. \ e_C( c_1 \mathsf{\wedge} c_2, x, y ) \teq (e_C( c_1, x, y ) \wedge e_C( c_2, x, y )) \\
\forall xy. \ e_S( \mathsf{y}, x, y ) \teq y & \forall c_1 c_2 xy. \ e_C( c_1 \mathsf{\vee} c_2, x, y ) \teq (e_C( c_1, x, y ) \vee e_C( c_2, x, y )) \\
\multicolumn{2}{l}{\forall s_1 s_2 xy. \ e_S( s_1 \mathsf{+} s_2, x, y ) \teq e_S( s_1, x, y ) + e_S( s_2, x, y )}  \\
\multicolumn{2}{l}{\forall c_1 s_1 s_2 xy. \ e_S( \mathsf{ite}( c_1, s_1, s_2 ), x, y ) \teq ite( e_C( c_1, x, y ), e_S( s_1, x, y ), e_S( s_2, x, y ) )}  \\
\multicolumn{2}{l}{\forall c_1 xy. \ e_C( \mathsf{\neg} c_1, x, y ) \teq \neg e_C( c_1, x, y )} \\
\end{array}\]
These evaluation operators define an interpreter for programs (terms of sort $S$ and $C$) given inputs $x$ and $y$.
The interpretation of a term $e_S( g, x, y )$ can be determined for any constant $g$, $x$, and $y$ using quantifier instantiation, 
where the number of instantiations required for doing so is (at most) the term size of $g$.

With these operators, our property $P_0$ can be restated as:
\[\begin{array}{l@{\hspace{1em}}l}
P := \lambda gxy. \ e_S( g, x, y ) \geq x \wedge e_S( g, x, y ) \geq y \wedge ( e_S( g, x, y ) \teq x \vee e_S( g, x, y ) \teq y )
\end{array}\]

When asked whether there exists an $f$ that satisfies this specification $P$,
we invoke the SMT solver to determine the satisfiability of $\mathcal{A}_S \cup \mathcal{A}_C \cup \forall g. \exists xy. \neg P( g, x, y )$,
where our background theory is the combination of linear arithmetic, datatypes, and uninterpreted functions.
Notice that instantiating the latter quantified formula with $\mathsf{ite( x \leq y, y, x )}$ for $g$ gives
us $\exists xy. \neg P( \mathsf{ite( x \leq y, y, x )}, x, y )$.
The solver will determine this is unsatisfiable
using the ground decision procedures in combination with quantifier instantiation for unfolding the evaluation of concrete programs for inputs.
We claim this suffices to show that $\mathsf{ite( x \leq y, y, x )}$ is a solution to the synthesis conjecture.
$\Box$
\end{example}

In remains to show how the SMT solver discovers that our quantified formula should be instantiated with $\mathsf{ite( x \leq y, y, x )}$.
Heuristic quantifier instantiation techniques, e.g. E-matching, 
are based around instantiating quantified formulas using terms already occurring in an input.
Clearly, since we are asking the solver to find an instantiation that represents a synthesized a term we have yet to see, 
these heuristics are likely ineffective for this purpose.
Our preliminary experiments confirm that heuristic instantiation techniques used by most modern SMT solvers are ineffective even for simple conjectures of the form mentioned in this document.
Instead, we present a specialized technique, which we refer to as \emph{counterexample-guided quantifier instantiation}, 
which can be used as an technique by the SMT solver to quickly converge on the instantiation that falsfies the synthesis conjecture.
The technique follows a popular scheme for synthesis known as counterexample-guided inductive synthesis,
which has been implemented in various systems such as Sketch~\cite{DBLP:conf/asplos/Solar-LezamaTBSS06}.

\sparagraph{Counterexample-Guided Quantifier Instantiation.}
Say we are given a negated synthesis conjecture $\psi := \forall g. \ \exists \vec i. \neg P( g, \vec i )$, where $g$ has sort $S$,
and an axiomization $\mathcal{A}$ defining how terms of sort $S$ evaluate.
To determine the satisfiability of $\mathcal{A} \cup \psi$ in some background theory $T$,
as with common approaches to SMT solving~\cite{nieuwenhuis2006solving}, our approach maintains a set of ground clauses $F$ (which in our case is initially empty).
Additionally, our approach makes use of two components:
\begin{itemize}
\item A fresh constant $e$ of sort $S$, the current candidate solution for $g$, and
\item A fresh predicate $G$, whose interpretation corresponds to possibility that $g$ has a solution.
\end{itemize}
Our procedure terminates either when $F$ is unsatisfiable, in which case we have found a solution,
or when all models of $F$ interpret $G$ as false, in which case we have shown that no solution exists.
It consists of two alternating steps, stated in the following:

\begin{enumerate}
\begin{framed}
\item If $F$ has a model $\M$ such that $\M( G ) = \top$, add $\neg P( \M( e ), \vec k )$ to $F$ for fresh $\vec k$ and go to step 2.  Otherwise, answer ``no solution".
\item If $F$ has a model $\M'$, add $G \Rightarrow P( e, \M'( \vec k ) )$ to $F$ and repeat step 1.  Otherwise, answer ``$\M( e )$ is a solution".
\end{framed}
\end{enumerate}

\ 

The above procedure, which we call counterexample-guided quantifier instantiation, has been implemented in the SMT solver \cvc~\cite{CVC4-CAV-11}.
Let us revisit the example of finding a function that computes the maximum of two integers $x$ and $y$.
One run of the steps of the above procedure are as follows (as computed by \cvc):
\[\begin{array}{l@{\hspace{1em}}c@{\hspace{1em}}l}
\hline
\text{Step} & \text{Model} & \text{Added Clause} \ 
\\
\hline
 1 & \{ e \mapsto \mathsf{x}, \ldots \} & \neg P( \mathsf{x}, x_1, y_1 ) \\
 2 & \{ x_1 \mapsto 0, y_1 \mapsto 1, \ldots \} & G \Rightarrow P( e, 0, 1 ) \\
 1 & \{ e \mapsto \mathsf{y}, \ldots \} & \neg P( \mathsf{y}, x_2, y_2 ) \\
 2 & \{ x_2 \mapsto 1, y_2 \mapsto 0, \ldots \} & G \Rightarrow P( e, 1, 0 ) \\
 1 & \{ e \mapsto \mathsf{1}, \ldots \} & \neg P( \mathsf{1}, x_3, y_3 ) \\
 2 & \{ x_3 \mapsto 2, y_3 \mapsto 0, \ldots \} & G \Rightarrow P( e, 2, 0 ) \\
 1 & \{ e \mapsto \mathsf{x+y}, \ldots \} & \neg P( \mathsf{x+y}, x_4, y_4 ) \\
 2 & \{ x_4 \mapsto 1, y_4 \mapsto 1, \ldots \} & G \Rightarrow P( e, 1, 1 ) \\
 1 & \{ e \mapsto \mathsf{ite( x \leq 1, 1, x ) }, \ldots \} & \neg P( \mathsf{ite( x \leq 1, 1, x )}, x_5, y_5 ) \\
 2 & \{ x_5 \mapsto 1, y_5 \mapsto 2, \ldots \} & G \Rightarrow P( e, 1, 2 ) \\
 1 & \{ e \mapsto \mathsf{ite( x \leq y, y, x ) }, \ldots \} & \neg P( \mathsf{ite( x \leq y, y, x )}, x_6, y_6 ) \\
 2 & \text{none} &  \\
\hline
\end{array}\]

In this run, notice that each model found for $e$ satisfies all values of counterexamples found for previous candidates.
After the sixth iteration of step 1, the procedure finds the candidate $\mathsf{ite( x \leq y, y, x )}$, for which no counterexample exists,
indicating that the procedure has found a solution for the synthesis conjecture.
At the moment, this problem can be solved in $\sim.5$ seconds in the latest development version of \cvc.

\sparagraph{Fairness.}
In our preliminary experience, a necessary technique for limiting the candidate programs is to consider smaller programs before larger ones.
Adapting techniques for finding finite models of minimal size~\cite{reynolds2013finite}, 
we use a strategy that searches for programs of size $1$ only after we have exhausted the search for programs of size $0$.
This can be accomplished in the DPLL(T) framework by introducing a splitting lemma of the form $( size( e ) \leq 0 \vee \neg size( e ) \leq 0 )$, and
asserting $size( e ) \leq 0$ as the first decision literal, where $size$ is a function mapping a datatype term to its term size 
(an integer corresponding to the number of non-nullary constructor applications in a term).
We do the same for $size( e ) \leq 1$ if and when $\neg size( e ) \leq 0$ becomes asserted.
The decision procedure for inductive datatypes in \cvc~\cite{BarST-JSAT-07} has been extended in our implementation to handle constraints involving $size$.

We state the following claims about our procedure here.

\begin{claim}
Using the aforementioned fairness strategy, the procedure mentioned in this section has the following properties:
\begin{enumerate}
\item (Solution Soundness) When it answers ``$f$ is a solution", then $\forall \vec i. P( f, \vec i )$ holds,
\item (Refutation Soundness) When it answers ``no solution", then $\forall \vec i. P( f, \vec i )$ does not hold for any $f$, and
\item (Solution Completeness) If the satisfiability problem for $P( e, \vec k )$ is decidable, and there exists an $f$ such that $\forall \vec i. P( f, \vec i )$ holds, it answers ``$g$ is a solution" for some $g$.
\end{enumerate}
\end{claim}

\subsection{General Synthesis for Single-Invocation Properties}
\label{sec:gen-synthesis}

Consider the case where no syntax is provided as input, and we are asked to find a function $f$ of sort $S_1 \times \ldots \times S_n \rightarrow S_r$
satisfying some universal property $\forall \vec i. P( f, \vec i )$, where all instances of $f$ occur as the term $f( \vec i )$.
We refer to such $P$ as a \emph{single-invocation property}.
Approaches for axiomatizations over such properties have been studied in~\cite{jacobs2011towards}.
We may rephrase synthesis conjectures for a single-invocation property $P$ as:
\begin{eqnarray} \label{eqn:syn_conj_no_syntax}
\forall \vec i. \exists g. Q( g, \vec i )
\end{eqnarray}
where $g$ is a variable of sort $S_r$.  
In contrast to the conjecture in~\eqref{eqn:syn_conj}, notice that the quantification on the function to synthesize has been pushed downwards.
Finding a model for this formula amounts to finding a skolem function of sort $S_1 \times \ldots \times S_n \rightarrow S_r$ for $g$.
This section describes a general approach for determining the satisfiability of formulas of this form.

If $Q( g, \vec i )$ resides in a particular fragment of first-order logic, 
say linear arithmetic, then determining the satisfiability of the above constraint can be accomplished using a method for quantifier elimination~\cite{monniaux2010quantifier, bjornerQE2010}.
Such cases have been examined in the context of software synthesis~\cite{KuncakETAL12SoftwareSynthesisProcedures}.
Alternatively, we may again explore an instantiation-based approach for establishing the unsatisfiability of the negated form of this conjecture:
\begin{eqnarray} \label{eqn:neg_syn_conj_no_syntax}
\exists \vec i. \forall g. \neg Q( g, \vec i )
\end{eqnarray}
The existence of a finite set of ground instantiations to show the formula~\eqref{eqn:neg_syn_conj_no_syntax} 
is unsatisfiable suffices to show the existence of a solution to our conjecture.
Moreover, when this is the case, solutions for $g$ may be constructed due to the following observation:
\begin{remark}
Say that $\neg Q( t_1, \vec k ), \ldots, \neg  Q( t_n, \vec k ) \models_T \bot$ for fresh constants $\vec k$.
Then, $\ell := \lambda \vec k. \ ite( Q( t_1, \vec k ), t_1, \ldots ite( Q( t_{n-1}, \vec k ), t_{n-1}, t_n ) \ldots )$ is a solution for $g$
in $\forall \vec i. \exists g. Q( g, \vec i )$.
\end{remark}
\begin{proof}
Given an arbitrary set of ground terms $\vec u$ of the same sort as $\vec i$ and model $\M$, we show that $\M \models Q( \ell( \vec u ), \vec u )$.
Let $\sigma$ be the substitution $\{ \vec k \mapsto \vec u \}$.
Consider the case that $\M \models Q( t_i \sigma, \vec u )$ for some (least) $i \in \{ 1, \ldots, n-1 \}$.
Then, $\M( \ell( \vec u ) ) = t_i \sigma$, and thus $\M \models Q( \ell( \vec u ), \vec u )$. 
If no such $i$ exists, then $\M \models \neg Q( t_i \sigma, \vec u )$ for all $i = 1, \ldots, n-1$, and $\M( \ell( \vec u ) ) = t_n \sigma$.
By our assumption and since $\vec k$ is fresh,
we have that $\neg Q( t_1 \sigma, \vec u ), \ldots, \neg  Q( t_{n-1} \sigma, \vec u ) \models_T Q( t_n \sigma, \vec u )$, which is $Q( \ell( \vec u ), \vec u )$.
\eop
\end{proof}

\ 


\begin{example}
Let us revisit the example of a function computing the max of its inputs $x$ and $y$.
In the absence of syntatic restrictions, this can be phrased as the following,
where $g$, $x$, and $y$ are variables of sort $Int$:
\begin{eqnarray}
Q := \lambda gxy. g \geq x \wedge g \geq y \wedge ( g \teq x \vee g \teq y )
\end{eqnarray}
Our negated synthesis conjecture is then $\exists xy. \ \forall g. \ \neg Q( g, x, y )$, which after skolemization is $\forall g. \ \neg Q( g, k_1, k_2 )$ for fresh constants $k_1$ and $k_2$.
When asked to determine the satisfiability of $\forall g. \ \neg Q( g, k_1, k_2 )$,
the SMT solver may, for instance, instantiate this formula with $k_1$ and $k_2$ for $g$, 
giving us $\neg Q( k_1, k_1, k_2 )$ and $\neg Q( k_2, k_1, k_2 )$ which together are unsatisfiable in the theory of linear arithmetic.
By the aforementioned remark,
this tells us that $\lambda k_1 k_2. \ ite( Q( k_1, k_1, k_2 ), k_1, k_2 )$ is a solution for $g$,
which is $\lambda k_1 k_2. \ ite( k_1 \geq k_1 \wedge k_1 \geq k_2 \wedge ( k_1 \teq k_1 \vee k_1 \teq k_2 ), k_1, k_2 )$
and simplifies to $\lambda k_1 k_2. \ ite( k_1 \geq k_2, k_1, k_2 )$.
$\Box$
\end{example}

It remains to be shown how the solver determines the instantiations $k_1$ and $k_2$ for $g$ in $\forall g. \neg Q( g, k_1, k_2 )$.
The procedure described in the previous section can be modified as follows.
We again maintain current set of clauses $F$, and introduce a fresh constant $e$ of the same sort as $g$, and guard predicate $G$.
To begin, skolemize the outermost quantifier of our conjecture, giving us the constraint $\forall g. \neg Q( g, \vec k )$ for fresh $\vec k$,
and add the clause $G \Rightarrow Q( e, \vec k )$ to $F$.
Our approach then consists of iterations of the following step, where we write $L( t_1, \ldots, t_n )$ as shorthand for 
$\lambda \vec k. \ ite( Q( t_1, \vec k ), t_1, \ldots ite( Q( t_{n-1}, \vec k ), t_{n-1}, t_n ) \ldots )$:
\begin{enumerate}
\begin{framed}
\item If $\neg Q( t_1, \vec k ), \ldots, \neg  Q( t_n, \vec k ) \subseteq F$ is unsat, answer ``$L( t_1, \ldots, t_n )$ is a solution".  If $F$ has a model $\M$ such that $\M( G ) = \top$, add $\neg Q( t, \vec k )$ to $F$ for some term $t$ where $\M(t) = \M(e)$, and repeat.  Otherwise, answer ``no solution".
\end{framed}
\end{enumerate}

The construction of term $t$ in this loop intentionally underspecified.
A na{\"i}ve choice for $t$ is the constant in our signature whose interpretation in a standard model is $\M( e )$.
This choice amounts to testing whether points in the range of the function satisfy the specification.
More sophisticated choices for $t$ are a subject of current work.


\subsection{Syntax-Guided Synthesis for Single-Invocation Properties}

Consider the case when both (1) our syntax $S$ for solutions contains the constructor $\mathsf{ite} : C \times S \times S \rightarrow S$ for some inductive datatype $C$,
and (2) the property we wish to synthesize is single-invocation and can be expressed as a term of sort $C$.
For instance, the property from Example~\ref{ex:max-sygus} can be phrased as:
\begin{eqnarray} \label{eqn:gen-sygus}
R:= \lambda g k_1 k_2. \ e_C( g \mathsf{ \geq x \wedge } ( g \mathsf{ \geq y \wedge } ( g \mathsf{ \teq x \vee } g \mathsf{ \teq y } ) ), k_1, k_2 )
\end{eqnarray}
where $g$ has sort $S$ and $k_1$ and $k_2$ have sort $Int$.
The method in Section~\ref{sec:gen-synthesis} is applicable to the conjecture $\exists xy. \ \forall g. \ \neg R( g, x, y )$ since it emits solutions meeting our syntactic requirements.
This has the advantage over the approach in Section~\ref{sec:syntax-guided} in that it only needs to synthesize the outputs of a solution, and not conditions in $\mathsf{ite}$-terms.
Our implementation in \cvc is capable of handling conjectures of this form, 
where we limit our choice of $t$ in the procedure described in the previous section to be the constant term $\M( e )$.
Assuming $k_1$ and $k_2$ are skolem constants for $x$ and $y$,
the run for this example is the following:

\[\begin{array}{l@{\hspace{1em}}c@{\hspace{1em}}l}
\hline
\text{Model} & \text{Choice of } t & \text{Added Clause} \ 
\\
\hline
\{ e \mapsto \mathsf{x}, \ldots \} & \mathsf{x} & \neg R( \mathsf{x}, k_1, k_2 ) \\
\{ e \mapsto \mathsf{y}, \ldots \} & \mathsf{y} & \neg R( \mathsf{y}, k_2, k_2 ) \\
\text{none} &  \\
\hline
\end{array}\]

This indicates that, for instance, $\mathsf{ ite( x \geq x \wedge ( x \geq y \wedge ( x \teq x \vee x \teq y ) ), x, y ) }$
is a solution for $g$, which subsequently could be simplified to $\mathsf{ ite( x \geq y, x, y ) }$.
Notice the method described in this section terminates after two iterations (in \lt .05 seconds)
as opposed to the method mentioned in Section~\ref{sec:syntax-guided}, which terminated after six iterations,
leading to a tenfold decrease in runtime for this example.




{
\bibliographystyle{abbrv}
\bibliography{main}

\begin{thebibliography}{10}

\bibitem{6679385}
R.~Alur, R.~Bodik, G.~Juniwal, M.~Martin, M.~Raghothaman, S.~Seshia, R.~Singh,
  A.~Solar-Lezama, E.~Torlak, and A.~Udupa.
\newblock Syntax-guided synthesis.
\newblock In {\em Formal Methods in Computer-Aided Design (FMCAD), 2013}, pages
  1--17, Oct 2013.

\bibitem{CVC4-CAV-11}
C.~Barrett, C.~Conway, M.~Deters, L.~Hadarean, D.~Jovanovic, T.~King,
  A.~Reynolds, and C.~Tinelli.
\newblock {CVC4}.
\newblock In {\em Proceedings of CAV'11}, volume 6806 of {\em LNCS}, pages
  171--177. Springer, 2011.

\bibitem{BarST-JSAT-07}
C.~Barrett, I.~Shikanian, and C.~Tinelli.
\newblock An abstract decision procedure for satisfiability in the theory of
  inductive data types.
\newblock {\em Journal on Satisfiability, Boolean Modeling and Computation},
  3:21--46, 2007.

\bibitem{beyene2014constraint}
T.~Beyene, S.~Chaudhuri, C.~Popeea, and A.~Rybalchenko.
\newblock A constraint-based approach to solving games on infinite graphs.
\newblock In {\em Proceedings of the 41st annual ACM SIGPLAN-SIGACT symposium
  on Principles of programming languages}, pages 221--234. ACM, 2014.

\bibitem{beyene2013solving}
T.~A. Beyene, C.~Popeea, and A.~Rybalchenko.
\newblock Solving existentially quantified horn clauses.
\newblock In {\em Computer Aided Verification}, pages 869--882. Springer Berlin
  Heidelberg, 2013.

\bibitem{DBLP:conf/sas/BjornerMR13}
N.~Bj{\o}rner, K.~L. McMillan, and A.~Rybalchenko.
\newblock On solving universally quantified {H}orn clauses.
\newblock In {\em SAS}, pages 105--125, 2013.

\bibitem{bjornerQE2010}
N.~Bjørner.
\newblock Linear quantifier elimination as an abstract decision procedure.
\newblock In J.~Giesl and R.~Hähnle, editors, {\em Automated Reasoning},
  volume 6173 of {\em Lecture Notes in Computer Science}, pages 316--330.
  Springer Berlin Heidelberg, 2010.

\bibitem{DBLP:conf/cade/MouraB07}
L.~de~Moura and N.~Bj{\o}rner.
\newblock Efficient {E-Matching} for {SMT} solvers.
\newblock In {\em CADE, July 17-20, 2007, Proceedings}, volume 4603 of {\em
  Lecture Notes in Computer Science}, pages 183--198. Springer, 2007.

\bibitem{Detlefs03simplify:a}
D.~Detlefs, G.~Nelson, and J.~B. Saxe.
\newblock Simplify: A theorem prover for program checking.
\newblock Technical report, J. ACM, 2003.

\bibitem{GeDeM-CAV-09}
Y.~Ge and L.~de~Moura.
\newblock Complete instantiation for quantified formulas in satisfiability
  modulo theories.
\newblock In {\em Proceedings of CAV'09}, volume 5643 of {\em LNCS}, pages
  306--320. Springer, 2009.

\bibitem{jacobs2011towards}
S.~Jacobs and V.~Kuncak.
\newblock Towards complete reasoning about axiomatic specifications.
\newblock In {\em Verification, Model Checking, And Abstract Interpretation},
  pages 278--293. Springer Berlin Heidelberg, 2011.

\bibitem{KuncakETAL12SoftwareSynthesisProcedures}
V.~Kuncak, M.~Mayer, R.~Piskac, and P.~Suter.
\newblock Software synthesis procedures.
\newblock {\em Communications of the ACM}, 2012.

\bibitem{monniaux2010quantifier}
D.~Monniaux.
\newblock Quantifier elimination by lazy model enumeration.
\newblock In {\em Computer Aided Verification}, pages 585--599. Springer Berlin
  Heidelberg, 2010.

\bibitem{nieuwenhuis2006solving}
R.~Nieuwenhuis, A.~Oliveras, and C.~Tinelli.
\newblock Solving sat and sat modulo theories: From an abstract
  davis--putnam--logemann--loveland procedure to dpll (t).
\newblock {\em Journal of the ACM (JACM)}, 53(6):937--977, 2006.

\bibitem{reynolds14quant_fmcad}
A.~Reynolds, C.~Tinelli, and L.~D. Moura.
\newblock Finding conflicting instances of quantified formulas in {SMT}.
\newblock In {\em Formal Methods in Computer-Aided Design (FMCAD)}, 2014.

\bibitem{reynolds2013finite}
A.~J. Reynolds.
\newblock {\em Finite Model Finding in Satisfiability Modulo Theories}.
\newblock PhD thesis, The University of Iowa, 2013.

\bibitem{DBLP:conf/asplos/Solar-LezamaTBSS06}
A.~Solar-Lezama, L.~Tancau, R.~Bod\'{\i}k, S.~A. Seshia, and V.~A. Saraswat.
\newblock Combinatorial sketching for finite programs.
\newblock In {\em ASPLOS}, pages 404--415, 2006.

\end{thebibliography}
}

\end{document}